\documentclass[11pt]{article}
\usepackage{pstricks}
\usepackage{bbm}
\usepackage{amsmath,amssymb,amsthm}
\usepackage{amsbsy}
\usepackage{epsfig}
\usepackage{cite}

\theoremstyle{definition}

\numberwithin{equation}{section}
\numberwithin{theorem}{section}
\newcommand{\comment}[1]{}

\begin{document}{}

\title{On Madelung Systems in Nonlinear Optics. \\ A Reciprocal Invariance}

\author{Colin Rogers$^\dag$ and Boris Malomed$^\ddag$\\[5pt]
$^\dag$School of Mathematics and Statistics, The University of New South Wales, \\Sydney, NSW2052, Australia\\
\texttt{c.rogers@unsw.edu.au} \\[5pt]
$^\ddag$The Iby and Aladar Fleischman Faculty of Engineering,\\ Tel Aviv University, Israel\\
\texttt{malomed@post.tau.ac.il}}

\maketitle

\begin{abstract}
The role of the de Broglie-Bohm potential, originally established as central to Bohmian quantum mechanics, is examined for two canonical Madelung systems in nonlinear optics. In a seminal case, a Madelung system derived by Wagner \textit{et al} via the paraxial approximation and in which the de Broglie-Bohm potential is present, is shown to admit a multi-parameter class of what are here introduced as `q-gaussons'. In the limit as the Tsallis parameter $q\rightarrow1$, the q-gaussons are shown to lead to standard gausson solitons as admitted by logarithmic nonlinear Schr\"odinger equation encapsulating the Madelung system. The q-gaussons are obtained for optical media with dual power-law refractive index.

In the second case, a Madelung system originally derived via an eikonal approximation in the context of laser beam propagation and in which the de Broglie Bohm term is neglected, is shown to admit invariance under a novel class of two-parameter class of reciprocal transformations. Model optical laws analogous to the celebrated K\'arm\'an-Tsien law of classical gasdynamics are introduced.
\end{abstract}

\section{Introduction}

The occurrence of de Broglie-Bohm-type terms $\nabla^2|\Psi|/|\Psi|$ in Madelung systems arising in nonlinear optics was early documented in the pioneering work of Wagner \textit{et al} \cite{wwhhjm68} on the application of the paraxial approximation in the context of the self-trapping of optical beams. This seminal optics system can be encapsulated in a generalised nonlinear Schr\"odinger equation which likewise involves a de Broglie-Bohm potential. Tatarinova and Garcia in a series of papers \cite{ltmg07,ltmg08,ltmg11} have subsequently investigated a Madelung system in nonlinear optics in which an eikonal approximation was adopted so that the de Broglie-Bohm term therein is neglected. Boundary value problems for this reduced Madelung system were investigated via a hodograph transformation for a variety of nonlinear model optical laws.

Here, it is established that a dual power-law nonlinear Schr\"odinger equation incorporating a de-Broglie Bohm potential admits what here are termed q-gaussons. The corresponding wave packet representations have the feature structure with an amplitude of the q-exponential type as  originally introduced by Tsallis in a statistical mechanics context (see, e.g., Gell-Mann and Tsallis \cite{mmct04}). The q-gaussons are here shown to be embedded in a more general class of exact solutions which includes, in particular, the Helmholz optical solitons of \cite{jcgmpp,jcgmpp07,jcgmpp10}. Importantly, it is shown that, the presence of the de Broglie-Bohm term in the Madelung system associated with the dual power-law NLS equation is key to the existence of this class of solutions.

In the case of the reduced Madelung system of \cite{ltmg07,ltmg08,ltmg11} with de Broglie-Bohm term neglected, invariance is established under a two-parameter $(\delta,\epsilon)$-class of reciprocal transformations. This allows the analytic results of \cite{ltmg07,ltmg08,ltmg11} to be extended to novel $(\delta,\epsilon)$-dependent classes of nonlinear optical model laws. An integrable reduction of the Madelung system of \cite{ltmg07,ltmg08,ltmg11} is obtained via a reciprocal transformation for nonlinear optics models analogous to the celebrated K\'arm\'an-Tsien pressure-density law of classical gasdynamics.

\section{A Madelung System in Nonlinear Optics. De Broglie-Bohm Potential}

In a pioneering paper of Wagner \textit{et al} \cite{wwhhjm68}, Maxwell's equations
\begin{equation} \label{b1}
\nabla \times \mathbf{E} = - \mu \frac{\partial \mathbf{H}}{\partial t} \ , \quad \nabla \times \mathbf{H} = \frac{\partial \mathbf{D}}{\partial t} \ , \end{equation}
with nonlinear constitutive law
\begin{equation} \label{b2}
\mathbf{D} = \epsilon_L \mathbf{E} + \chi (\mathbf{E}) \mathbf{E} \end{equation}
which relates the displacement flux density vector $\mathbf{D}$ and the electric field $\mathbf{E}$ was used to derive a Madelung system modelling the self-trapping of optical beams in the paraxial approximation. This system adopted the form
\begin{equation} \label{b3}
\kappa \frac{\partial E^2_0}{\partial z} + \nabla_T (E^2_0\nabla_T\Phi) =0 \ , \end{equation}
\begin{equation} \label{b4}
2\kappa \frac{\partial \Phi}{\partial z} + |\nabla_T \Phi|^2 - \frac{\nabla^2_T E_0}{E_0} - \kappa^2 \chi (E^2_0)/\epsilon_L = 0 \ , \end{equation}
wherein $E_0$ is a slowly varying electric field amplitude, $\Phi$ the phase, and  cylindrical symmetry was assumed. In the above, $\nabla_T$ denotes the gradient perpendicular to the direction of propagation, while
\begin{equation} \label{b5}
\begin{array}{c} \mathbf{E} = E_0 \exp i \; \Phi \cdot \exp [-i (\omega t - \kappa z)] , \\[0.1in]
(\kappa^2 = \epsilon_L \mu \; \omega^2) \ . \end{array} \end{equation}

On setting $\rho=E^2_0$ and use of the scaling $t=z/\kappa$, the coupled nonlinear system (\ref{b3})$-$(\ref{b4}) becomes
\begin{equation} \label{b6}
\frac{\partial \rho}{\partial t} + (\nabla_T \rho) \cdot \nabla_T \Phi + \rho  \nabla^2_T \Phi = 0 \ , \end{equation}
\begin{equation} \label{b7}
\frac{\partial \Phi}{\partial t} + \frac{1}{2} |\nabla_T \Phi|^2 - \frac{1}{2} \frac{\nabla^2_T \rho^{1/2}}{\rho^{1/2}} - \left(\frac{\kappa^2}{2\epsilon_L}\right) \chi (\rho) = 0 \ . \end{equation}
Introduction of the Madelung transformation
\begin{equation} \label{b8}
\Psi = \rho^{1/2} \exp \left(\frac{i \Phi}{2}\right) \end{equation}
shows that the system (\ref{b6})$-$(\ref{b7}) may be encapsulated in an NLS-type equation
\begin{equation} \label{b9}
\begin{array}{c} i \ \dfrac{\partial \Psi}{\partial t} + \nabla^2 \Psi + \sigma \chi (|\Psi|^2) \Psi = s \left(\dfrac{\nabla^2 |\Psi|}{|\Psi|}\right) \Psi \\[0.2in]
(\sigma = \kappa^2/4\epsilon \ , \ s = 3/4) \end{array} \end{equation}
incorporating a de Broglie-Bohm potential term $\nabla^2|\Psi|/|\Psi|$ which  has its origin in de Broglie-Bohm quantum theory \cite{db52,lb65}. If $s<1$ as is the case for the NLS equation \eqref{b9} which encapsulates the Madelung system \eqref{b6}$-$\eqref{b7}, the de Broglie-Bohm term can be removed by an appropriate transformation (see e.g. \cite{crbmkcha10,cr14}).

In general, the NLS equation \eqref{b9} is not integrable. However, in 1+1-dimensions when $s<1$ and $\chi(|\Psi|^2)\sim|\Psi|^2$ corresponding to a standard Kerr medium reduction, the equation is tantamount to the standard integrable cubic NLS equation. In that case, it inherits the established characteristic properties of that solitonic equation such as being amenable to the inverse scattering procedure and admitting invariance under a B\"acklund transformation with its concomitant geometric properties and nonlinear superposition principle (see e.g. \cite{mapc92,crws82,crws02,crws98} and literature cited therein). If $s>1$ as in the cold plasma context of \cite{jlopcrws07,crpc18}, the NLS equation \eqref{b9} with $\chi(|\Psi|^2\sim|\Psi|^2$ embodies a basic two-component integrable system contained with the AKNS hierarchy of \cite{madkanhs73}. This `resonant' NLS equation with $s>1$ can admit novel solitonic fission or fusion phenomena \cite{opjlcr08}. It likewise admits an auto-B\"acklund transformation and associated nonlinear superposition principle \cite{jlopcrws07}.

It remains of importance to isolate other model $\chi(|\Psi|^2)$ relations and constraints on the parameter $s$ such that NLS equations such as \eqref{b9} admit exact wave packet representations of physical interest. In this connection, it is recorded below that for $\chi(|\Psi|^2)$ incorporating a logarithmic nonlinearity, \eqref{b9} admits 3+1-dimensional gaussons.

\vspace{0.2in}\centerline{\textbf{Gaussons}}

\vspace{0.1in} In \cite{thdaml10}, Hansen \textit{et al} considered a logarithmic saturable medium wherein the change $\delta n$ in the refractive index $n(\mathcal{I})$ was assumed to depend on intensity according to
\begin{equation} \label{b10}
\delta n(\mathcal{I}) \sim \ln (\mathcal{I}/\mathcal{I}_s) \end{equation}
where $\mathcal{I}_s$ is a constant which characterises the saturation intensity. Therein, a logarithmic NLS equation was set down for the slowly varying wave field amplitude and a perturbation approach was adopted to analyse coherent `soliton' interaction. Nonlinear media with logarithmic dependence of the refractive index on the wave intensity has been investigated in a number of works (see e.g. \cite{dctcrj97,wkdeob00,hbasmstsmsdc03,crha11}).

The nonlinear Schr\"odinger equation \eqref{b9} with model law $\sigma\chi(|\Psi|^2)=\alpha\ln|\Psi|+\beta$, namely
\begin{equation} \label{b11}
i \frac{\partial \Psi}{\partial t} + \nabla^2 \Psi + (\alpha \ln |\Psi| + \beta) \Psi = s \left( \frac{\nabla^2|\Psi|}{|\Psi|}\right) \Psi \end{equation}
admits gausson-type solutions
\begin{equation} \label{b12}
\Psi = Ne^{i \{\omega t - (k_1x+k_2y+k_3z+k_4)\}} \cdot e^{-a[(x-x_0-\nu_1t)^2+(y-y_0-\nu_2t)^2+(z-z_0-\nu_3t)^2]} \ . \end{equation}
of the type introduced in \cite{ibjm78}, in the absence of the de-Broglie Bohm potential. Thus, insertion of the wave packet representation \eqref{b12} into \eqref{b11} establishes that such gausson solutions exist for the Madelung system \eqref{b6}$-$\eqref{b7} with logarithmic law $\sigma\chi=\alpha\ln|\Psi|+\beta$, provided the parameters are such that
\begin{equation} \label{b13}
\begin{array}{c} \alpha = 4a (1 - s) \ , \quad \beta = \Sigma^3_{i=1} k^2_i - 6a s + \omega \ , \\[0.1in]
\nu_i + 2k_i = 0 \ , \qquad i = 1, 2, 3 \ . \end{array} \end{equation}
This class of gausson solutions is such that the initial conditions
\begin{equation} \label{b14}
|\Psi|_{t=o} = e^{-a[(x-x_0)^2+(y-y_0)^2+(z-z_0)^2]} \ , \quad |\nabla|\Psi||_{t=o} = (k^2_1+k^2_2+k^2_3)^{1/2} \end{equation}
hold. Importantly, it is seen that such gaussons are precluded in the case $s=1$. This will be seen to be the case for the 1+1-dimensional Madelung system of \cite{ltmg07,ltmg08,ltmg11}.

It is remarked that a modulated version of the logarithmic NLS equation \eqref{b12} incorporating a de Broglie-Bohm potential has been recently shown in \cite{cr2014} to admit integrable subsystems of Ermakov-Ray-Reid type \cite{jr80,jrjr80,crchjr93,crws96}. Two-component coupled nonlinear systems of the latter kind have been shown to arise in a range of nonlinear optics and other physical contexts (see e.g. \cite{crbmkcha10,crha10,crbmha12} and literature cited therein).

\section{q-Gaussons}

The notions of a q-logarithmic and its inverse q-exponential function namely
\begin{equation} \label{c1}
\left. \begin{array}{l} \log_q x = \dfrac{x^{1-q}-1}{1-q} \ , \\[0.2in]
\exp_q x = [ 1 + (1-q) x ]^{1/(1-q)} \ , \end{array}
\begin{array}{c} q \neq 1 \end{array} \right\} \end{equation}
were originally introduced by Tsallis in a statistical mechanics context and have a broad spectrum of physical applications (see \cite{ct09,jn11} and literature cited therein; these functions should not be confused
with the Jackson's q-exponential functions, which were introduced much earlier as basic functions of
the ``q-calculus" \cite{Jackson}). The classical logarithmic and exponential functions are retrieved in the limit as the Tsallis parameter $q\rightarrow1$.

In anisotropic gasdynamics \cite{crtr14} and magnetogasdynamics incorporating rotation \cite{crws14}, 2+1-dimensional elliptic vortex motions have recently been isolated corresponding to q-Gaussian density distributions
\begin{equation} \label{c2}
\begin{array}{c} \rho = \omega(t) \exp_q [\ -\mathbf{x}^T \mathbf{E}(t) \mathbf{x} + \mathbb{C}\ ] \ , \\[0.1in]
\mathbf{x} = \left( \begin{array}{c} x \\ y \end{array} \right) \end{array} \end{equation}
and with q-dependent gas laws. In the magnetogasdynamic setting of \cite{crws14}, such motions were obtained for q-dependent constitutive laws relating the pressure $p$, density $\rho$ and entropy $S$, namely
\begin{equation} \label{c3}
p = \rho^\gamma \Sigma(S) \ , \quad \Sigma \sim \left[ \exp_q \left(- \dfrac{S}{\mathcal{A}_0} \right) \right]^{2-q-\gamma} \end{equation}
corresponding to a particular Prim gas \cite{crwswh02}. It is noted that \eqref{c3} represents a `q-deformation' of the classical polytropic gas law
\begin{equation} \label{c4}
p = \rho^\gamma \Sigma(S) \ , \quad \Sigma = \mathcal{R} e^{(\gamma-1)S/\mathcal{R}} \end{equation}
where $\gamma$ is the adiabatic index.

Here, 1+1-dimensional dual power law NLS equations incorporating a de Broglie-Bohm potential are considered of the type
\begin{equation} \label{c5}
i \ \frac{\partial \Psi}{\partial t} + \delta \ \frac{\partial^2 \Psi}{\partial t^2} + \frac{\partial^2 \Psi}{\partial x^2} + ( \lambda + \mu |\Psi|^{q-1} + \nu |\Psi|^{2(q-1)} ) \Psi = s \left(\frac{|\Psi|_{xx}}{|\Psi|}\right) \Psi \end{equation}
and wave packet representations incorporating what are here termed `q-gaussons', with
\begin{equation} \label{c6}
\Psi = \mathcal{N} e^{-iEt} e^{il(x-kt)} \exp_q [\ -a(x-kt)^2\ ] = \mathcal{N} e^{i(-Et+l(x-kt))} [\ 1 - a(x - kt)^2 (1-q)\ ]^{\frac{1}{1-q}} \ , \end{equation}
are isolated. It is remarked the occurrence of dual power-law NLS equations with $\delta=0$ and $s=0$ is well-documented and arises in the study of nonlinear optical materials with refractive indices of the type \cite{khmk95,rmvaykjl96,dpvayk96,naaarg99,ab04,absk06}
\begin{equation} \label{c7}
n (|\mathbf{E}|) = n_\sigma |\mathbf{E}|^\sigma + n_{2\sigma} |\mathbf{E}|^{2\sigma} \ . \end{equation}

If we proceed with the representation
\begin{equation} \label{c8}
\Psi = \mathcal{N} e^{i(-Et+l(x-kt))} \Phi (x - kt) \ , \end{equation}
then insertion into \eqref{c5} yields
\begin{equation} \label{c9}
k = 2l + 2k \delta (E + kl) \end{equation}
together with
\begin{equation} \label{c10}
( \delta k^2 + 1 - s) \Phi'' + [\ E + kl - \delta (E + kl)^2 - l^2 + \lambda + \mu \mathcal{N}^{q-1} \Phi^{q-1} + \nu \mathcal{N}^{2(q-1)} \Phi^{2(q-1)}\ ] \Phi = 0 \ . \end{equation}
It is noted that the latter kind of nonlinear equation arises elsewhere in optical waveguide theory in \cite{asjl83}.

In view of \eqref{c6}, exact solutions of \eqref{c10} are sought with
\begin{equation} \label{c11}
\Phi = \phi(\zeta)^{\tfrac1{1-q}} \ , \quad \zeta = x - kt \ , \end{equation}
so that
\begin{equation} \label{c12}
\begin{array}{l} \left[ \dfrac{q}{(1-q)^2} \phi^{\frac{2q-1}{1-q}} \phi^{'2} + \dfrac{1}{1-q} \phi^{\frac{q}{1-q}} \phi'' \right] ( \delta k^2 + 1 - s) \\[0.2in]
\quad + \left[ E + kl - \delta (E + kl)^2 - l^2 + \lambda + \mu \mathcal{N}^{q-1} \phi^{-1} + \nu \mathcal{N}^{2(q-1)} \phi^{-2} \right] \phi^{\frac{1}{1-q}} = 0 \ . \end{array} \end{equation}
where the prime denotes a derivative with respect to $\zeta$. The latter admits the class of exact solutions with
\begin{equation} \label{c13}
\phi'' + \Lambda \phi = \mathbb{K} \end{equation}
so that
\begin{equation} \label{c14}
\begin{array}{c}\phi^{'2} + \Lambda \phi^2 = 2 \mathbb{K} \phi + \mathbb{L} \ , \\ [0.1in]
\mathbb{K}, \mathbb{L} \in \mathbb{R} \end{array} \end{equation}
and with a triad of relations between the parameters, namely
\begin{equation} \label{c15}
- \frac{\Lambda}{(1-q)^2} ( \delta k^2 + 1 - s) + E + kl - \delta (E + kl)^2 - l^2 + \lambda = 0 \ , \end{equation}
\begin{equation} \label{c16}
\mathbb{K} \left( \frac{1+q}{(1-q)^2} \right) ( \delta k^2 + 1 - s) + \mu \mathcal{N}^{q-1} = 0 \ , \end{equation}
\begin{equation} \label{c17}
\frac{\mathbb{L}q}{(1-q)^2} ( \delta k^2 + 1 - s) + \nu \mathcal{N}^{2(q-1)} = 0 \ . \end{equation}

Accordingly, dual power-law NLS equations \eqref{c5} incorporating a de Broglie-Bohm potential admit the class of solutions \eqref{c11}, modulo the relations \eqref{c15}$-$\eqref{c17}, wherein
\begin{equation} \label{c18}
\phi = \left\{ \begin{array}{l} c_1\phi_1 + c_2\phi_2 + \mathbb{K}/\Lambda \ , \quad \Lambda \neq 0 \ , \\[0.1in]
\mathbb{K} (\zeta^2/2) + \mathbb{M}\zeta + \mathbb{N} \ , \quad  \Lambda = 0 \end{array} \right. \end{equation}
with $\phi_1,\phi_2$ linearly independent solutions of
\begin{equation} \label{c19}
\phi'' + \Lambda \phi = 0 \ . \end{equation}
It is noted that there is the additional important requirement
\begin{equation} \label{c20}
\delta k^2 + 1 - s \neq 0 \ . \end{equation}
in \eqref{c10}. This condition is seen to hold, in particular, for the Madelung system obtained via the paraxial approximation by Wagner \textit{et al} \cite{wwhhjm68}, wherein $\delta=0,s=3/4$.

\vspace{0.2in}\centerline{\textbf{q-gaussons}}

\vspace{0.1in} The class of q-gaussons \eqref{c6} is admitted as the specialisation $\Lambda=0$ in \eqref{c18} with
\begin{equation} \label{c21}
\mathbb{K} = 2a(q - 1) \ , \quad \mathbb{M} = 0 \ , \mathbb{N} = 1 \ . \end{equation}
In the limit $q\rightarrow1$ it incorporates the gausson solutions to the 1+1-dimensional version \eqref{b11} of the logarithmic NLS equation with a de Broglie-Bohm quantum potential term.

\vspace{0.2in}\centerline{\textbf{Helmholz Solitons}}

\vspace{0.1in} Christian \textit{et al} \cite{jcgmpp,jcgmpp07,jcgmpp10} have investigated the phenomenon of `Helmholz solitons' which arise as exact solutions of a dual power-law NLS equation descriptive of the propagation of optical beams in certain plane wave guides. This nonlinear Schr\"odinger equation corresponds with appropriate scaling and the specialisation of \eqref{c5} with $\delta\neq0$ but $s=0$ so that de Broglie-Bohm term is absent. In this case, the condition \eqref{c20} requires that $\delta\neq-1/k^2$ so that $\delta>0$. The Helmholz solitons of \cite{jcgmpp,jcgmpp07,jcgmpp10} are retrieved in the case $s=0$ as the particular class in \eqref{c18} with $\Lambda\neq0$ and $\phi_1\sim \coth\zeta,\ \phi_2=0$. Importantly, it is here seen that such Helmholz solitons likewise exist for the NLS equation \eqref{c5} incorporating a de Broglie-Bohm potential as in \cite{wwhhjm68}.

\section{Invariance of a Madelung System under a Class of Reciprocal Transformations}

Tatarinova and Garcia in a series of papers \cite{ltmg07,ltmg08,ltmg11} have investigated certain boundary value problems for a Madelung system in nonlinear optics obtained via an eikonal approximation in which the de Broglie-Bohm term is neglected. The respective 1+1-dimensional Madelung system was derived as a reduction of the NLS equation \cite{rb03}
\begin{equation} \label{d1}
i \ \frac{\partial \mathcal{E}}{\partial z} + \frac{1}{2k_0} \nabla^2_\perp \mathcal{E} + k_0 n(|\mathcal{E}|^2) \mathcal{E} = 0 \end{equation}
to model laser pulse propagation in a medium with nonlinear refractive index $n(|\mathcal{E}|^2)$. Here, $\mathcal{E}=\sqrt{\mathcal{I}}\exp(ik_0S)$ is the electric field, $\mathcal{I}$ is the electric field intensity: $z$ is the propagation length and $k_0=n_0\omega_0/c$ is a wave number, where $\omega_0$ is the carrier frequency of the laser radiation and $c$ is the speed of light.

The eikonal approximation of \cite{ltmg07,ltmg08,ltmg11} resulted in the Madelung system
\begin{equation} \label{d2}
\frac{\partial \mathcal{I}}{\partial z} + \frac{\partial}{\partial x} (\mathcal{I} v) = 0 \ , \end{equation}
\begin{equation} \label{d3}
\frac{\partial v}{\partial z} + v \frac{\partial v}{\partial x} - \frac{\partial}{\partial x} n(\mathcal{I}) = 0 \end{equation}
where $v=S_x$. The latter system may be encapsulated in an NLS equation
\begin{equation} \label{d4}
i \Psi_z + \Psi_{xx} + \frac{1}{2} n(\mathcal{I}) \Psi = \frac{s|\Psi|_{xx}\Psi}{|\Psi|} \end{equation}
with $\Psi=\mathcal{I}^{1/2}e^{iS/2}$ and $s=1$. Thus, importantly, since $\delta=0$ in this case (c.f. \eqref{c5}), it is seen that $\delta k^2+1-s=0$ so that the requirement \eqref{c20} precludes the existence of q-gaussons and Helmholz solitons in the above Madelung system \eqref{d2}$-$\eqref{d3} of \cite{ltmg07,ltmg08,ltmg11}. However, the latter like its classical gasdynamic counterpart, admits linearization via a hodograph transformation involving interchange of the dependent and independent variables. In \cite{ltmg07,ltmg08,ltmg11}, a class of boundary value problems consisting of the $(v,\mathcal{I})$-system \eqref{d2}$-$\eqref{d3} augmented by the side conditions
\begin{equation} \label{d5}
v|_{z=0} = 0 \ , \quad \mathcal{I}|_{z=0} = \mathcal{I}_0 \exp (-x^2/\omega^2_{in}) \ , \quad \omega_{in} = \omega_0/\sqrt{2} \end{equation}
was thereby analysed for various nonlinear optical terms $n(\mathcal{I})$.

Here, it is shown that the Madelung system \eqref{d2}$-$\eqref{d3} is privileged in that it admits invariance under a two-parameter class of reciprocal transformations.

\vspace{0.2in}\newpage\centerline{\textbf{A Reciprocal Invariance}}

\vspace{0.1in} Reciprocal transformations have wide application in nonlinear continuum mechanics, notably in gasdynamics, magnetogasdynamics and the exact solution of nonlinear moving boundary problems \cite{cr68,cr69,cr85,cr15,afcrws05}. In modern soliton theory, they have importance in the linkage of inverse scattering schemes and the canonical integrable equations contained therein (see e.g. \cite{crws02,crpw84}). Here, a novel application of a reciprocal transformation in the context of nonlinear optics is presented.

\vspace{0.2in}\centerline{\textbf{A Reciprocal Invariance}}
\vspace{0.05in}\centerline{\textbf{Theorem}}

\vspace{0.1in} The 1+1-dimensional Madelung system (\ref{d2})$-$(\ref{d3}) is invariant under the two-parameter $(\delta,\epsilon)$-class of relations
\begin{equation} \label{d6}
\left. \begin{array}{c} \mathcal{I}^* = \dfrac{\epsilon \ \mathcal{I}}{1+\delta\mathcal{I}} \ , \quad v^* = v \\[0.2in]
dn^* = (1 + \delta\mathcal{I}) dn \ , \quad dx^* =  (1 + \delta\mathcal{I}) dx - \delta\mathcal{I} \ v \ dz \ , \quad dz^* = dz \end{array} \right\} \mathbb{R}^* \ . \end{equation}
\hspace{5.5in}$\square$

Thus, under $\mathbb{R}^*$
\begin{equation*}
\mathcal{I}^* dx^* - \mathcal{I}^* v^* dz^* = \epsilon (\ \mathcal{I} \ dx - \mathcal{I} \ v \ dz\ ) \ , \end{equation*}
so that
\begin{equation} \label{d7}
\frac{\partial \mathcal{I}^*}{\partial z^*} + \frac{\partial}{\partial x^*} (\mathcal{I}^* v^*) = 0 \ . \end{equation}
In addition, the Madelung system admits the conservation law
\begin{equation} \label{d8}
\frac{\partial}{\partial z} (\mathcal{I} \ v) + \frac{\partial}{\partial x}\ (p \  \mathcal{I} + \mathcal{I} \ v^2) = 0 \end{equation}
where
\begin{equation} \label{d9}
dp = - \mathcal{I} d ( n(\mathcal{I})) \ . \end{equation}
Thus,
\begin{equation} \label{d10}
\mathcal{I}^* v^* dx^* - (p^* + \mathcal{I}^* v^{*2}\ ) dz^* = \frac{\epsilon \mathcal{I} v}{1+\delta \mathcal{I}} [\ (1+\delta\mathcal{I}) dx - \delta\mathcal{I} \ v \ dz\ ] - \left( \frac{\epsilon v^2}{1+\delta \mathcal{I}} + p^* \right) dz \end{equation}
whence with $p^*=\epsilon p$ so that
\begin{equation} \label{d11}
dn^* = (1 + \delta \mathcal{I}) dn \ , \end{equation}
\eqref{d10} yields
\begin{equation*}
\mathcal{I}^* v^* dx^* - (p^* + \mathcal{I}^* v^{*2}\ ) dz^* = \epsilon [\  \mathcal{I} \ v \ dx - (p + \mathcal{I} v^2) dz\ ] \end{equation*}
so that
\begin{equation} \label{d12}
\frac{\partial}{\partial z^*}\ ( \mathcal{I} v^*) + \frac{\partial}{\partial x^*}\ (p^* (\mathcal{I}^*) + \mathcal{I}^* v^{*2}) = 0 \ . \end{equation}

Invariance of the Madelung system \eqref{d2}$-$\eqref{d3} under the two-parameter $(\delta,\epsilon)$-class of reciprocal transformations $\mathbb{R}^*$ given by \eqref{d6} has accordingly been established. This novel reciprocal invariance permits analytic results established in the series of papers \cite{ltmg07,ltmg08,ltmg11} concerning the boundary value problem \eqref{d5} for the Madelung system \eqref{d2}$-$\eqref{d3} to be extended to a more general class of model $(\delta,\epsilon)$-dependent $n(\mathcal{I})$-relations and associated boundary value problems. Thus, in particular, in the case of the standard nonlinear Kerr medium with $n(\mathcal{I})=\nu\mathcal{I}$, it is seen that $\mathcal{I}^*dn^*=\epsilon\mathcal{I} \ \nu \ d\mathcal{I}$ and one obtains the associated reciprocal Kerr class of $(\delta,\epsilon)$-dependent  $n^*(\mathcal{I})^*$ model relations given by
\begin{equation} \label{d13}
\boxed{n^* = \frac{\nu}{2\delta \epsilon^2 (1 - (\delta/\epsilon) \mathcal{I}^*)^2} + \text{const} \ . } \end{equation}
It is recalled that model $n(\mathcal{I})$ expressions of the type
\begin{equation} \label{d14}
n(\mathcal{I}) = \frac{\alpha}{1+\sigma \mathcal{I}} + \xi \ , \end{equation}
where in \cite{vkykvs98}
\begin{equation} \label{d15}
\xi = - \alpha = \frac{1}{\sigma} \left( \frac{1-\sigma}{1+\sigma} \right) \end{equation}
have been widely used to model nonlinear saturable media (see e.g. \cite{absk06} and literature cited therein). The reciprocal model nonlinear Kerr law \eqref{d13} admits saturation phenomena if $\delta>0,\epsilon<0$.

Under the reciprocal transformations $\mathbb{R}^*$ given by \eqref{d6}, the class of boundary value problems determined by the Madelung system \eqref{d2}$-$\eqref{d3} subject to the conditions \eqref{d5} is embedded in a more general class valid for the reciprocal $n^*(\mathcal{I}^*; \delta; \epsilon)$ model relations. Corresponding to a solution $\{\mathcal{I}(x,z), v(x,z)\}$ of the original Madelung system is the class of solutions of the reciprocal Madelung system
\begin{equation} \label{d16}
\frac{\partial \mathcal{I}^*}{\partial z^*} + \frac{\partial}{\partial x^*} (\mathcal{I}^* \ v^*) = 0 \ , \quad \frac{\partial v^*}{\partial z^*}+ v^* \frac{\partial v^*}{\partial x^*} - \frac{\partial}{\partial x^*} n^* (\mathcal{I}^*; \delta, \epsilon) = 0 \end{equation}
given parametrically in terms of $x,z$ by the reciprocal relations
\begin{equation} \label{d17}
\mathcal{I}^* = \frac{\epsilon \ \mathcal{I}}{1+\delta \mathcal{I}} \ , \quad v^* = v \ , \end{equation}
where
\begin{equation} \label{d18}
x^* = x + \delta X \end{equation}
with $X(x,z)$ the Lagrangian variable determined by
\begin{equation} \label{d19}
dX = \mathcal{I} \ dx - \mathcal{I} \ v \ dz \ . \end{equation}

\section{Canonical Reduction}

Here, it is noted that the nonlinear optics Madelung system \eqref{d2}$-$\eqref{d3} with de Broglie-Bohm term absent, under the correspondence $\mathcal{I}\rightarrow\rho, \ dn\rightarrow -\rho^{-1}dp$ produces the classical 1+1-dimensional gasdynamic system with barotropic $p(\rho)$-relation relating the pressure $p$ and density $\rho$. Whereas the latter system and the corresponding Madelung system \eqref{d2}$-$\eqref{d3} can, via a hodograph transformation, be reduced to a linear system, the latter involves terms dependent on the constitutive $(p,\rho)$-law in gasdynamics or $n(\mathcal{I})$-term in the present nonlinear optics context.

The celebrated two-parameter model K\'arm\'an-Tsien $(p,\rho)$-pressure-gas relation
\begin{equation} \label{e1}
p = \frac{A}{\rho} + B \end{equation}
was originally introduced to approximate, via reduction of the classical hodograph equations to the Cauchy-Riemann system, the plane subsonic motion of an adiabatic gas with $p\sim\rho^\gamma$ (see e.g. \cite{ws54} and work cited therein). Loewner \cite{cl50,cl52}, subsequently applied a novel class of matrix B\"acklund transformations to construct model multi-parameter $(p,\rho)$-laws in plane gasdynamics for which the hodograph system can be reduced to appropriate canonical forms in subsonic, transonic and supersonic flow r\'egimes. These B\"acklund transformations may be applied `mutatis mutandis' to construct model constitutive laws in both nonlinear elastostatics and elastodynamics \cite{crws10,crws2010}. Nonlinear superposition principles and unexpected solitonic connections have been established therein. Moreover, a re-interpretation and generalisation of the Loewner class of infinitesimal B\"acklund transformations of \cite{cl52} was shown in \cite{bkcr91,bkcr93} to provide novel linear representations for a wide class of 2+1-dimensional integrable systems which may be parametrised in terms of a triad of eigenfunctions. Importantly, the B\"acklund transformations of \cite{cl50,cl52} may be readily adapted to the nonlinear optics context of \cite{ltmg07,ltmg08,ltmg11} using the correspondence $\mathcal{I}\rightarrow\rho, \ dn\rightarrow -\rho^{-1}dp$ to construct model $n(\mathcal{I})$-terms for which the Madelung system \eqref{d2}$-$\eqref{d3} may be reduced to tractable canonical form.

\vspace{0.2in}\centerline{\textbf{A Reciprocal Reduction}}

\vspace{0.1in} Here, to conclude, a class of model $n(\mathcal{I})$-relations is recorded which corresponds to the K\'arm\'an-Tsien model law \eqref{e1} of gasdynamics and which allows reduction via a reciprocal transformation of the Madelung system \eqref{d2}$-$\eqref{d3} to a pair of classical 1+1-dimensional wave equations.

Thus, on introduction of the reciprocal transformation
\begin{equation} \label{e2}
\left. \begin{array}{c} \mathcal{I}' = \dfrac{1}{\mathcal{I}} \ , \quad v' = v \\[0.2in]
dx' =  \mathcal{I} \ dx - \mathcal{I} \ v \ dz \ , \quad dz' = dz \end{array} \right\} \ \ \mathbb{R}' \end{equation}
the Madelung system \eqref{d2}$-$\eqref{d3} adopts the Lagrangian form
\begin{equation} \label{e3}
\frac{\partial \mathcal{I}'}{\partial z'} = \frac{\partial v'}{\partial x'} \ , \quad - \frac{1}{\mathcal{I}^{'3}} \ (dn/d\mathcal{I}) \ \frac{\partial \mathcal{I}'}{\partial x'} = \frac{\partial v'}{\partial z'} \ . \end{equation}
Accordingly, with model $n(\mathcal{I})$-laws
\begin{equation} \label{e4}
n(\mathcal{I}) = \frac{c^2}{2\mathcal{I}^2} + \text{const} \end{equation}
the system \eqref{e3} reduces to the canonical form
\begin{equation} \label{e5}
\frac{\partial \mathcal{I}'}{\partial z'} = \frac{\partial v'}{\partial x'} \ , \quad \frac{1}{c^2} \ \frac{\partial \mathcal{I}'}{\partial x'} = \frac{\partial v'}{\partial z'} \end{equation}
which encodes on elimination in turn of $v'$ and $\mathcal{I}'$, classical 1+1-dimensional wave equations for $\mathcal{I}'$ and $q'$. Thus,
\begin{equation} \label{e6}
\mathcal{I} = \frac{1}{\phi(x'+cz)+\psi(x'-cz)} \ , \quad v = c [\ \phi(x' +cz) - \psi(x'-cz)\ ] \end{equation}
where $\phi,\psi$ have arbitrary dependence on their indicated arguments. The expressions in \eqref{e6} together with the reciprocal relation for $dx'$ in \eqref{e2} which provides the integrable expression
\begin{equation} \label{e7}
dx = [\ \phi(x' +cz) + \psi(x'-cz)\ ] dx' + c [\ \phi(x' +cz) - \psi(x'-cz)\ ] dz \ , \end{equation}
yield a parametric representation in temrs of $x',z$ for the solution of the Madelung system \eqref{d2}$-$\eqref{d3} corresponding to the class of model optical $n(\mathcal{I})$-relations \eqref{e4}. It is remarked that the latter are distinctive in that they remain invariant in form under the class of reciprocal transformations $R^*$. Moreover, the relations $\mathcal{I}\rightarrow\rho, \ dn\rightarrow -\rho^{-1}dp$ may be used to show that the class of model $n(\mathcal{I})$-terms \eqref{e4} correspond to a class of K\'arm\'an-Tsien laws of gasdynamics.

\section{Conclusion}

Here, two kinds of Madelung system which arise in nonlinear optics have been investigated and shown, in turn, to admit q-gausson solutions and invariance under a novel two-parameter class of reciprocal transformations. In addition, a canonical reduction has been presented for a class of nonlinear refractive $n(\mathcal{I})$-relations analogous to the K\'arm\'an-Tsien law of classical gasdynamics. The latter connection suggests, more generally, the potential application in nonlinear optics of Loewner-type B\"acklund transformations to construct model $n(\mathcal{I})$-relations for which the hodograph system as investigated in the work of Tatarinova and Garcia \cite{ltmg07,ltmg08,ltmg11} may be reduced systematically to tractable canonical form.

\def\bitem{\vspace{-0.2cm}\bibitem}
\def\fit#1{\textit{\frenchspacing#1}}

\end{document}